\documentclass[letterpaper, 10 pt, conference,usenames,dvipsnames,final]{ieeeconf}  

\IEEEoverridecommandlockouts                              
\let\labelindent\relax

\title{\LARGE Dynamics Concentration of Large-Scale Tightly-Connected Networks}
\author{Hancheng Min and Enrique Mallada
\thanks{H. Min and E. Mallada are with the Department of Electrical and Computer Engineering, Johns Hopkins University, Baltimore, MD 21218, USA{\tt\small \{hanchmin, mallada\}@jhu.edu}.}
\thanks{The work was supported by ARO through contract W911NF-17-1-0092, US DoE EERE award DE-EE0008006, and NSF through grants CNS 1544771, EPCN 1711188, AMPS 1736448, and CAREER 1752362.}
}

\usepackage{mysty}
\usepackage{bbold,amsmath,amsthm,amsfonts,amssymb}
\usepackage{subfiles}
\allowdisplaybreaks
\usepackage{environ}
\usepackage{ifthen}
\usepackage{cite}
\usepackage{accents}
\usepackage{balance}
\usepackage{verbatim}

\newboolean{archive}
\setboolean{archive}{true}

\usepackage{tikz}
\usetikzlibrary{shapes, arrows}
\usepackage{enumitem}
\tikzstyle{block} = [draw, rectangle, minimum height=3.5em, minimum width=3.5em]
\tikzstyle{sum} = [draw, circle, node distance=1cm]
\tikzstyle{input} = [coordinate] \tikzstyle{output} = [coordinate]
\tikzstyle{tmp} = [coordinate]

\def\thesubsubsection{\arabic{section}}
\newtheorem{thm}{Theorem}[subsubsection]
\newtheorem{lem}[thm]{Lemma}
\newtheorem{col}[thm]{Corollary}

\newtheorem*{dfn}{Definition}
\theoremstyle{remark}
\newtheorem*{rem}{Remark}

\begin{document}
\thispagestyle{empty}
\pagestyle{empty}
\maketitle
\begin{abstract}
    The ability to achieve coordinated behavior --engineered or emergent-- on networked systems has attracted widespread interest over several fields. This has led to remarkable advances on the development of a theoretical understanding of the conditions under which agents within a network can reach agreement (consensus) or develop coordinated behaviors such as synchronization. However, fewer advances have been made toward explaining another commonly observed phenomena in tightly-connected networks systems:  output responses of nodes in the networks are almost identical to each other despite heterogeneity in their individual dynamics.
    In this paper, we leverage tools from high-dimensional probability to provide an initial answer to this phenomena. More precisely, we show that for linear networks of nodal random transfer functions, as the network size and connectivity grows, every node in the network follows the same response to an input or disturbance --irrespectively of the source of this input. We term this behavior as \emph{dynamics concentration} since it stems from the fact that the network transfer matrix uniformly converges in probability, i.e., it concentrates, to a unique dynamic response determined by the distribution of the random transfer function of each node.
    We further discuss the implications of our analysis in the context of model reduction and robustness, and provide numerical evidence that similar phenomena occur in small deterministic networks over a properly defined frequency band.
\end{abstract}

\section{Introduction}\label{sec:intro}

    Coordinated behavior in network systems has been a popular subject of research in many fields, such as physics~\cite{Bressloff1999}, chemistry~\cite{Kiss2002}, social sciences~\cite{DeGroot1974}, and biology~\cite{Mirollo1990}. Within engineering, coordination is essential for proper operation of many networked systems including power networks~\cite{jpm2017cdc,Paganini2019}, data and sensor networks~\cite{mmhzt2015ton,m2014phd-thesis}, and autonomous transportation~\cite{Olfati-Saber2007,Jadbabaie2003988,Bamieh2012,Sepulchre2008706}.
    While there exist many expressions of this behavior in network systems, two forms of coordination have particularly received thorough attention by the control community: Consensus and synchronization.

    Consensus~\cite{DeGroot1974, Olfati-Saber2007, Jadbabaie2003988, Bamieh2012,Olfati-Saber20041520, Ghaedsharaf2019}, on one hand, refers to the ability of the network nodes to asymptotically reach a common value over some quantities of interest.
    Many extensions of this problem include the study of robustness and performance of consensus networks in the presence of noise~\cite{Jadbabaie2003988,Bamieh2012}, time-delay~\cite{Olfati-Saber20041520, Ghaedsharaf2019}, and switching graph topology~\cite{Ghaedsharaf2019}.
    Synchronization~\cite{Mirollo1990,mmhzt2015ton,m2014phd-thesis,Sepulchre2008706,Nair2008661,Kim2011200,Wieland2011}, on the other hand, refers to the ability of network nodes to follow a commonly defined trajectory. Although for nonlinear systems synchronization is a structurally stable phenomenon, in the linear case~\cite{Nair2008661, Sepulchre2008706, Kim2011200, Wieland2011}, synchronization requires the existence of a common internal model that acts as a virtual leader~\cite{Kim2011200,Wieland2011}.


    A less studied phenomenon, that is empirically observed, is a coherent response within large-scale tightly-connected networks in which every node identically reacts to perturbations, irrespectively from the nature or location of the disturbance. For example, in tightly-connected power networks, generator dynamics across the network tend to react coherently to system disturbances~\cite{Chow2013,Ariff2013,Guggilam2018}. However, while in the case of swing dynamics, reduced order models provide good approximations~\cite{Chow2013}, for generators with heterogeneous turbine time constants, a good low order approximation is difficult to find~\cite{Guggilam2018}. Among other contributions, this work aims to explain this difference.

    In this paper, we introduce a new framework to analyze the aggregated dynamics of large networks. We consider a network consisting of heterogeneous linear nodes interconnected through a weighted graph Laplacian matrix, with the node dynamics represented by random transfer functions.
    We show that, whenever the algebraic connectivity of the graph is polynomial in the network size $n$, the transfer matrix of the network converges in probability as $n$ grows to infinity to a common scalar transfer function spanning the consensus subspace. Notably, the resulting scalar transfer function is deterministic and can be determined by the harmonic expectation of the individual nodal dynamics. We term this behavior \emph{dynamics concentration}, due to the mathematical principle that explains this phenomenon, i.e., concentration of measure~\cite{Ledoux2001}.

    The implications of our results are manifold. Firstly, it extends the notions of consensus and synchronization to scenarios in which coherent behavior can be achieved even in the presence of disturbances that are arbitrary in source or shape. Secondly, unlike output synchronization that requires the existence of a common internal model within each node, dynamics concentration can be achieved despite heterogeneity on the individual dynamics or lack of a common internal model. Thirdly, due to the stochastic nature of our analysis, many networks that a priori may look quite different in composition and topology, exhibit exactly the same behavior. Finally, the analysis further provides a principled methodology to compute the concentrated dynamics, which as we will show later may not always be represented by a reduced order model.


    The rest of the paper is organized as follows. In Section \ref{sec:prelim}, we introduce some technical preliminaries such as the notion of sub-Gaussian random variables and their concentration inequalities and use this formalism to formulate our problem statement. In Section \ref{sec:stoc_convergence}, the condition for uniform stochastic convergence of the transfer matrix over compact sets is given. In Section \ref{sec:numerical}, we provide an application of our analysis to the problem of characterizing reduced models for power networks. At last, we conclude this paper with more discussions on the implications of this result.

\emph{Notation:}~For a vector $x$, $\|x\|=\sqrt{x^Tx}$ denotes the $2$-norm of $x$, and for a matrix $A$, $\|A\|$ denotes the spectral norm and $\underaccent{\bar}{\sigma}(A)$ denotes the least singular value of $A$. We let $I_n$ denote the identity matrix of order $n$, $V^*$ denote the conjugate transpose of matrix $V$, $\one$ denote $[1,\cdots,1]^T $ with proper dimensions, and  $[n]$ denote the set $\{1,2,\cdots,n\}$. Also, we write complex numbers as $a+jb$, where $j=\sqrt{-1}$.

For function ordering, we write $f(n)=o(g(n))$ if $\forall m>0$, $\exists n_0>0$ s.t. $|f(n)|<m g(n)$ for $n>n_0$. We write $f(n)=\Omega (g(n))$ if $\exists m>0$, $\exists n_0>0$ s.t. $f(n)\geq m g(n)$ for $n>n_0$.

\section{Preliminaries}\label{sec:prelim}

    
    \subsection{Sub-Gaussian random variables}
    Firstly, we let $(\Omega, \mathcal{F},\prob)$ be the probability space, where $\Omega$ is the sample space, $\mathcal{F}=2^\Omega$ is the event set and $\prob$ is a probability measure on $\mathcal{F}$. Wherever random variables are introduced, we assume that those random variables are properly defined in this probability space.
    
    Throughout the paper, we work on a special family of random variables: the sub-Gaussian random variables. One way to define such random variables is the following.
    \begin{dfn}[Sub-Gaussian random variables]
		A random variable $X$ is a sub-Gaussian random variable if $\forall t>0$:
		\begin{equation*}
		    \prob(|X|\geq t)\leq 2\exp\lp-ct^2\rp
		\end{equation*}
		for some $c>0$.
	\end{dfn}
	Such random variables exhibit an exponentially decaying tail probability, which gives good concentration results when summing over sub-Gaussian random variables. 
	
	A widely used bound on the tail probability of the sum is given by: 
    
\begin{lem}[Hoeffding's inequality\cite{Vershynin2018}]

\label{hoeff_ineq}

Let $X_1,\cdots,X_n$ be independent, mean zero, sub-gaussian random variables, and let $a=[a_1,\cdots,a_n]^T\in\ecud^n$. Then $\forall t> 0$, the following holds:
\begin{displaymath}
    \prob\lp \lv\sum_{i=1}^{n}a_iX_i\rv \geq t\rp\leq 2\exp\lp -\frac{ct^2}{\|a\|^2}\rp
\end{displaymath}
for some $c>0$.
\end{lem}

    We also provide a direct application of Hoeffding's inequality to the random complex numbers: 
    
\begin{lem}

\label{lem_hoeff_compl}

Let $(X_i,Y_i),\ i=1,\cdots,n$ be i.i.d. samples from a joint distribution such that $X_i,Y_i$ are sub-gaussian random variables, and let $a=[a_1,\cdots,a_n]^T\in\compl^n$. Then $\forall t> 0$, the following holds:
\begin{displaymath}
    \prob\lp \lv\sum_{i=1}^{n}a_i(X_i+jY_i)\rv \geq t\rp\leq 8\exp\lp -\frac{ct^2}{\|a\|^2}\rp
\end{displaymath}
for some $c>0$.
\end{lem}

    \ifthenelse{\boolean{archive}}{The proof is shown in appendix.}{Due to space constraints, we refer to~\cite{min2019} for the proof.}
    
    
    \subsection{Problem Statement}
    Consider a network consisting of $n$ nodes, indexed by $i\in[n]$ with the block diagram structure in Fig.1. We use $G(s)=\mathrm{diag}\{g_i(s)\}$, with $g_i(s),\ i\in[n]$, to represent the dynamics of the nodes. $L(s)$ is a generalized dynamic Laplacian matrix of the graph describing the network interconnection. More precisely, $L(s)$ is defined as transfer matrix such that $\forall s_0\in\compl$, $L(s_0)$ is a complex normal matrix with simple eigenvalue $\lambda=0$. In most applications, $L(s)=L\in\mathbb{R}^{n\times n}$ for some fixed graph Laplacian $L$. 
    
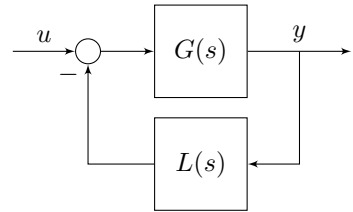
\begin{figure}[ht]
	\centering
	\begin{tikzpicture}[auto, node distance=1.5cm,>=latex']
	\node [input, name=input] {};
	\node [sum, right of=input] (sum) {};
	\node [block, right of=sum] (plant) {$G(s)$};
	\node [output, right of=plant, node distance=2cm] (output) {};
	\node [block, below of=plant] (laplacian) {$L(s)$};
	\draw [draw,->] (input) -- node {$u$} (sum);
	\draw [->] (sum) -- (plant);
	\draw [->] (plant) -- node [name=y]{$y$} (output);
	\draw [->] (y) |- (laplacian);
	\draw [->] (laplacian) -| node[pos=0.95]{$-$}(sum);
		
	\end{tikzpicture}
	\caption{Block Diagram of General Networked Dynamical Systems}\label{blk_p_n}
\end{figure}

    Many existing networks can be represented represented by this structure: For the first-order consensus\cite{Olfati-Saber2007}, the node dynamics is given by $g_i(s)=\frac{1}{s}$. For transportation networks, $g_i(s)$ are the vehicle dynamics. The Laplacian $L(s)=L$ is static for these two cases. For power networks\cite{Paganini2019}, $g_i(s)$ are the dynamics of the generators and $L(s)=\frac{1}{s}L_B$ where $L_B$ is a Laplacian matrix representing the sensitivity of power injection w.r.t. bus phase angles.
   
    The transfer matrix from $u$ to $y$ is given by:
    \begin{align*}
        T(s)&=\; (I_n+G(s)L(s))^{-1}G(s)\\
        &=\; (I_n+\mathrm{diag}\{g_i(s)\}L(s))^{-1}\mathrm{diag}\{g_i(s)\}
    \end{align*}

    Since L(s) is a normal complex matrix,
    \begin{equation}\label{laplcian_decomp}
        L(s)=V(s)\Lambda(s) V^*(s)\,,
    \end{equation} where $V(s)V^*(s)=V^*(s)V(s)=I_n$, and $\Lambda(s)=\mathrm{diag}\{\lambda_i(s)\}$ with $0=\lambda_1(s)<|\lambda_2(s)|\leq \cdots\leq |\lambda_n(s)|$. Using \eqref{laplcian_decomp}, we rewrite $T(s)$ as
    \begin{align}
        T(s)&=\; (I_n+\mathrm{diag}\{g_i(s)\}L(s))^{-1}\mathrm{diag}\{g_i(s)\}\nonumber\\
        &=\; (\mathrm{diag}\{g^{-1}_i(s)\}+L(s))^{-1}\nonumber\\
        &=\; (\mathrm{diag}\{g^{-1}_i(s)\}+V(s)\Lambda(s) V^*(s))^{-1}\nonumber\\
        &=\; V(s)(V^*(s)\mathrm{diag}\{g^{-1}_i(s)\}V(s)+\Lambda(s))^{-1}V^*(s)\label{eq1_sec3_subsec1}
    \end{align}
    When considering the convergence of the transfer matrix of networked dynamical systems, as the network size $n$ grows, it is intuitive to assume the node dynamics $g_i(s)$ are generated randomly according to a particular distribution and discuss the convergence in probability. In practice, the assumed distribution is modeled from physical parameters of the nodes. For example, we may assume that nodes dynamics are given by random rational transfer functions whose coefficients are random variables. 
    \begin{rem}
    For networks with deterministic nodes, notice that we can always assume the deterministic dynamics are random samples drawn from some unknown distributions, hence the empirical distribution is a reasonable approximation given some mild requirements on the statistics of the samples.
    \end{rem}
    We particularly focus on the convergence of $T(s)$ over a compact region $S\subset \compl$ where for $s_0\in S$, both the real and imaginary part of $g^{-1}_i(s_0)$ are sub-Gaussian random variables.
    Let
    \begin{align}
        \bar{g}(s)\!:&=\!\expc^{-1}[g_i^{-1}(s)]\!\label{eq2_sec3_subsec1}\\&=\![\expc Re(g_i^{-1}(s))\!+\!j\expc Im(g_i^{-1}(s))]^{-1}\nonumber
    \end{align}
    We will show that for networks with algebraic connectivity $\inf_{s\in S}|\lambda_2(s)|=\Omega(n^p)$ for some $p\in(0,1]$, as the network size $n$ increases, $T(s)$ converges uniformly to $\frac{1}{n}\bar{g}(s)\one\one^T$ in probability over $S$, i.e., $\forall\epsilon>0$
    \begin{equation*}
        \lim_{n\rightarrow\infty}\prob\lp \sup_{s\in S}\lV T(s)-\frac{1}{n}\bar{g}(s)\one\one^T\rV\geq \epsilon\rp=0
    \end{equation*}
    or using the following notation:
    \begin{equation*}
        \sup_{s\in S}\lV T(s)-\frac{1}{n}\bar{g}(s)\one\one^T\rV\xrightarrow{\mathcal{P}}0\quad \text{as}\ n\rightarrow\infty
    \end{equation*}
    Thus, since the output response of a stable system can be obtained by inverse Laplace transform on the imaginary axis,  for low frequency disturbances or input signals, having uniform convergence of $T(s)$ within a low frequency band on the imaginary axis, implies that, with high probability, the network output response is close to the output response of $\frac{1}{n}\bar{g}(s)\one\one^T$, i.e., the output responses of nodes in the networks are almost identical to each other, despite of the heterogeneity in their individual dynamics.

\section{Dynamics concentration of large-scale networks}\label{sec:stoc_convergence}
    
We term the uniform convergence of $T(s)$ over low frequency bands as \emph{dynamics concentration} because nodes with heterogeneous dynamics exhibit identical responses to low frequency disturbance, irrespectively of the source or shape of the disturbance. In this section, we firstly show point-wise convergence of $T(s)$ given the algebraic connectivity of the network is sufficiently large. Then we provide conditions for uniform convergence of $T(s)$ over a compact subset of the complex plane.
\subsection{Point-wise convergence in probability}
For convenience,  we assume at $s=s_0$, $g_i^{-1}(s_0)=X_i+jY_i$ where the $\{(X_i,Y_i)\}_{i=1}^{n}$ pairs are independent and $X_i, Y_i$ are sub-Gaussian random variables. Moreover, we let $L(s_0)=L$, with unitary decomposition:
\begin{equation}
    L =V\Lambda V^*\label{laplcian_decomp},\ \Lambda=\mathrm{diag}\{\lambda_i\}\,,
\end{equation}Then we have:
\begin{align}
    T(s_0)&=\;V(V^*\mathrm{diag}\{g_i^{-1}(s_0)\}V+\Lambda)^{-1}V^*\nonumber\\
    &=\;V(V^*\mathrm{diag}\{X_i+jY_i\}V+\Lambda)^{-1}V^*:=T
\end{align}
We also define $\mu:=\expc X_i+j\expc Y_i=\bar{g}^{-1}(s_0)$. Then the point-wise convergence in probability of $T(s)$ at $s=s_0$ is equivalent to
\begin{equation*}
    \lV T-\frac{1}{n}\mu^{-1}\one\one^T\rV\xrightarrow{\mathcal{P}}0\quad \text{as}\ n\rightarrow\infty\,,
\end{equation*}
which is simply the convergence of a random complex matrix in probability. To show convergence then, we firstly define
\begin{equation}
    H = V^*\mathrm{diag}\{X_i+jY_i\}V+\Lambda
\end{equation}
Since the first eigenvector of $L$ is $v_1=\frac{1}{\sqrt{n}}\one$, it follows that
\begin{align}
    \lV T-\frac{1}{n}\mu^{-1}\one\one^T\rV&=\; \|T-\mu^{-1}Ve_1e_1^TV^*\|\nonumber\\
    &=\; \lV V\lp H^{-1}-\mu^{-1}e_1e_1^T\rp V^*\rV\nonumber\\
    &=\; \lV H^{-1}-\mu^{-1}e_1e_1^T \rV\label{eq_norm_equiv}
\end{align}
where $e_1$ is the first column of identity matrix. The remaining is straightforward: we get an explicit form for $H^{-1}$ and bound the tail probability of $\lV H^{-1}-\mu^{-1}e_1e_1^T \rV$ by $o(1)$ terms.

To invert $H$, we let $\Tilde{X}_i=X_i-\expc X_i$, and $\Tilde{Y}_i=Y_i-\expc Y_i$, then rewrite $H$ as:
\begin{equation}
    H = V^*\mathrm{diag}\{\Tilde{X}_i+j\Tilde{Y}_i\}V+\mathrm{diag}\{\lambda_i+\mu\}
\end{equation}
The matrix $H$ can be regarded as a perturbed diagnal matrix $\mathrm{diag}\{\lambda_i+\mu\}$ with matrix $D:=V^*\mathrm{diag}\{\Tilde{X}_i+j\Tilde{Y}_i\}V$. The matrix $D$ is regarded as a perturbation in the sense of the following result:

\begin{lem}
\label{lem_concenration}
Let $(\Tilde{X}_i,\Tilde{Y}_i),\ i=1,\cdots,n$ be i.i.d. samples from a joint distribution such that $\Tilde{X}_i,\Tilde{Y}_i$ are mean zero, sub-gaussian random variables. Let $D=V^*\mathrm{diag}\{\Tilde{X}_i+j\Tilde{Y}_i\}V:=[d_{kl}]$, where $V=[v_1,\cdots,v_n]^T$ is defined in \eqref{laplcian_decomp}.\\ 
The following inequalities hold:
\begin{enumerate}
    \item $\forall \epsilon>0$
    \begin{equation}\label{lem_conc_eq1}
        \prob\lp |d_{11}|\geq \epsilon\rp\leq 8\exp\lp-c_1 n\epsilon^2\rp
    \end{equation}
    for some $c_1>0$.
    \item $\forall \epsilon>0$, $\forall p>0$
    \begin{subequations}
        \begin{equation}\label{lem_conc_eq2_sub1}
            \prob\lp \sum_{k=1}^n|d_{k1}|^2\geq \epsilon n^p\rp\leq \frac{c_2}{\epsilon n^p}
        \end{equation}
        \begin{equation}\label{lem_conc_eq2_sub2}
            \prob\lp \sum_{l=1}^n|d_{1l}|^2\geq \epsilon n^p\rp\leq \frac{c_2}{\epsilon n^p}
        \end{equation}
    \end{subequations}
    for some $c_2>0$.
    \item $\forall \epsilon>0$, $\forall p>0$
    \begin{equation}
        \prob\lp \|D\|\geq \epsilon n^p\rp\leq \frac{c_3\sqrt{\log n}}{\epsilon n^p}
    \end{equation}
    for some $c_3>0$.
\end{enumerate}
\end{lem}

\ifthenelse{\boolean{archive}}{The proof is shown in the appendix.}{For a proof, please refer to~\cite{min2019}.} Lemma \ref{lem_concenration} suggests that the perturbation matrix $D$ exhibits good concentration on its first row and column. We will exploit this property with a proper form of $H^{-1}$.

Define the following matrix:
\begin{equation}
    D_{(1)(1)}=\begin{bmatrix}
    d_{22}&\cdots&d_{2n}\\
    \vdots&\ddots&\vdots\\
    d_{n2}&\cdots&d_{nn}
    \end{bmatrix},\ \Lambda_{(1)}=\mathrm{diag}\{\lambda_i\}_{i=2}^n
\end{equation}
by removing the first row and column of $D$ and $\Lambda$. We then write $H$ as:
\begin{equation}\label{def_H}
    H=\left[\begin{array}{c|c}
    h_{11}&h_{12}^T\\ \hline
    h_{21}&H_{22}
    \end{array}\right],\ \begin{array}{l}
         h_{11}=d_{11}+\mu  \\
         h_{12}=[d_{12},\cdots,d_{1n}]^T\\
         h_{21}=[d_{21},\cdots,d_{n1}]^T\\
         H_{22}=D_{(1)(1)}+\Lambda_{(1)}+\mu I_{n-1}
    \end{array}
\end{equation}
Then, the inverse of $H$ is given by:
\begin{equation}\label{def_H_inv}
    \def\arraystretch{1.4}
    H^{-1}=\left[\begin{array}{c|c}
    \frac{1}{a}&-\frac{1}{a}h_{12}^TH_{22}^{-1}\\ \hline
    -\frac{1}{a}H_{22}^{-1}h_{21}&H_{22}^{-1}+\frac{1}{a}H_{22}^{-1}h_{21}h_{12}^TH_{22}^{-1}
    \end{array}\right]
\end{equation}
where $a=h_{11}-h_{12}^TH_{22}^{-1}h_{21}$. 

With this explicit form of $H^{-1}$, we bound the tail probability of $\|H^{-1}-\mu e_1e_1^T\|$ using the following lemma.

\begin{lem}

\label{lem_spec_norm_bd}
Assume $|\lambda_2|=\Omega(n^p)$ for some $p\in(0,1]$. For $H$ as defined in \eqref{def_H}, whenever $\mu\neq 0$, given $n$ large enough, then $\forall \epsilon>0$, we have:
\begin{align}
    &\; \prob\lp\lV H^{-1}-\mu^{-1}e_1e_1^T\rV\geq \epsilon\rp\nonumber\\
    \leq &\; 8\prob\lp|d_{11}|\geq C_1\rp+34\prob\lp\|D\|\geq C_2 n^p\rp\nonumber\\
    &\; \ +12\prob\lp \sum_{k=1}^n|d_{k1}|^2\geq C_3^2 n^{p}\rp+12\prob\lp\sum_{l=1}^n|d_{1l}|^2\geq C_3^2 n^{p}\rp\label{eq_lem_spec_norm_bd}
\end{align}
for some $C_1,C_2,C_3>0$ that depends on $\epsilon$.
\end{lem}

\ifthenelse{\boolean{archive}}{The proof is shown in the appendix.}{For proof, please refer to~\cite{min2019}.} As suggested in Lemma \ref{lem_concenration}, all terms on the right-hand side of \eqref{eq_lem_spec_norm_bd} are $o(1)$. By \eqref{eq_norm_equiv} and taking limit $n\rightarrow \infty$ on both sides of \eqref{eq_lem_spec_norm_bd}, we can then show the convergence in probability of $T$.

\begin{thm}

\label{thm_conv_spec_norm}

Let $(X_i,Y_i),\ i=1,\cdots,n$, be i.i.d. samples from a joint distribution such that $X_i,Y_i$ are sub-gaussian random variables, and denote $\mu:=\expc X_i+j\expc Y_i$. Consider a Laplacian matrix $L=V\Lambda V^*$, as defined in \eqref{laplcian_decomp}, with algebraic connectivity satisfying $|\lambda_2|= \Omega(n^p)$ for some $p\in(0,1]$. Then given
\begin{equation*}
    T= V(V^*\mathrm{diag}\{X_i+jY_i\}V+\Lambda)^{-1}V^*,
\end{equation*}
whenever $\mu\neq 0$, we have, $\forall \epsilon>0$,
\begin{equation*}
    \lim_{n\rightarrow +\infty}\prob\lp\lV T-\frac{1}{n}\mu^{-1}\one\one^T\rV\geq \epsilon\rp=0
\end{equation*}
\end{thm}

We can now apply the theorem to show the point-wise convergence of $T(s)$ at certain $s_0$.

\begin{col}\label{col_conv_spec_norm}
Let $T(s)$ and $\bar{g}(s)$ be defined as in \eqref{eq1_sec3_subsec1} and \eqref{eq2_sec3_subsec1}, respectively. Given $s=s_0$, consider the (possibly complex) network graph Laplacian $L(s_0)$ with algebraic connectivity satisfying $|\lambda_2(s_0)|=\Omega(n^p)$ for some $p\in(0,1]$.
Suppose that $g_i^{-1}(s_0)$ has both its real and imaginary part given by sub-gaussian random variables, and $s_0$ is not a pole of $\bar{g}(s)$. Then $\forall \epsilon>0$,
\begin{equation*}
    \lim_{n\rightarrow +\infty}\prob\lp\lV T(s_0)-\frac{1}{n}\bar{g}(s_0)\one\one^T\rV\geq \epsilon\rp=0\\
\end{equation*}
\end{col}

\ifthenelse{\boolean{archive}}{\subsection{Uniform convergence in probability}
For people who are familiar with real analysis, it is well-known that the uniform convergence of function sequence over a compact region not only requires point-wise convergence, but also needs the equicontinuity of the function sequence~\cite{Rudin1964}. Similarly, for stochastic uniform convergence of $T(s)$, we have the following theorem:

\begin{thm}

\label{thm_unifm_conv}
Let $T(s)$ and $\bar{g}(s)$ be defined as in \eqref{eq1_sec3_subsec1} and \eqref{eq2_sec3_subsec1}, respectively. Given $S$ a compact subset of $\compl$, and $\bar{g}(s)$ is uniform continuous on $S$, then $T(s)$ uniformly converges in probability to $\frac{1}{n}\bar{g}(s)\one\one^T$, i.e. $\forall \epsilon>0$
\begin{equation*}
    \lim_{n\rightarrow\infty}\prob\lp\sup_{s\in S}\lV T(s)-\frac{1}{n}\bar{g}(s)\one\one^T\rV\geq \epsilon\rp=0
\end{equation*}
if the following hold:
\begin{enumerate}
    \item (Point-wise convergence) $\forall s\in S$ and $\forall \epsilon>0$,
    \begin{equation*}
        \lim_{n\rightarrow\infty}\prob\lp\lV T(s)-\frac{1}{n}\bar{g}(s)\one\one^T\rV \geq \epsilon\rp=0
    \end{equation*}
    \item (Stochastic equicontinuity) $\forall\epsilon>0$,
    \begin{equation*}
        \lim_{\delta\rightarrow 0}\limsup_{n\rightarrow \infty}\prob\lp \sup_{\scriptscriptstyle\substack{|s_1-s_2|<\delta\\ s_1,s_2\in S}}\|T(s_1)-T(s_2)\|\geq \epsilon\rp=0
    \end{equation*}
\end{enumerate}
\end{thm}

\ifthenelse{\boolean{archive}}{We only provide a proof sketch here.}{} For details, please refer to~\cite{Newey1991}, where the uniform convergence of real-valued random function sequence is discussed.
\ifthenelse{\boolean{archive}}{
\begin{proof}[Sketch of proof]
Since $S$ is compact, for every $\delta>0$, there exists a finite cover $\{x\in\compl: |x-s_i|<\delta\}$, $i=1,\cdots,M$ with $s_i\in S,\ i=1,\cdots,M$.

One can show that, for every $\delta<0$, the following holds:
\begin{align}
    &\;\prob\lp\sup_{s\in S}\lV T(s)-\frac{1}{n}\bar{g}(s)\one\one^T\rV\geq \epsilon\rp\nonumber\\
    \leq &\; \prob\lp \sup_{\scriptscriptstyle \substack{|s_1-s_2|<\delta\\ s_1,s_2\in S}}\|T(s_1)-T(s_2)\|\geq \frac{\epsilon}{3}\rp\nonumber\\
    &\;\quad +\prob\lp \sup_{\scriptscriptstyle \substack{|s_1-s_2|<\delta\\ s_1,s_2\in S}}\|\bar{g}(s_1)-\bar{g}(s_2)\|\geq \frac{\epsilon}{3}\rp\nonumber\\
    &\;\quad +\sum_{i=1}^M\prob\lp \lV T(s_i)-\frac{1}{n}\bar{g}(s_i)\one\one^T\rV \geq \frac{\epsilon}{3}\rp\label{eq_temp_1}
\end{align}
For all $\eta>0$, choose $\delta$ small enough such that: 1) By uniform continuity of $\bar{g}(s)$, the second term is zero; 2) By stochastic equicontinuity of $T(s)$, 
\begin{equation}\label{eq_temp_2}
    \limsup_{n\rightarrow \infty}\prob\lp \sup_{\scriptscriptstyle \substack{|s_1-s_2|<\delta\\ s_1,s_2\in S}}\|T(s_1)-T(s_2)\|\geq \frac{\epsilon}{3}\rp<\frac{\eta}{2}
\end{equation}
Then for this $\delta$, choose $N>0$ such that for $n\geq N$, the first term is less than $\frac{\eta}{2}$, by \eqref{eq_temp_2}, and the third term is less than $\frac{\eta}{2}$, by point-wise convergence. Therefore we bound the left-hand side of \eqref{eq_temp_1} by arbitrary small $\eta$ for large $n$, which shows the uniform convergence.  
\end{proof}
}{}

The stochastic equicontinuity, as its name suggests, is the "stochastic version" of equicontinuity in the real analysis. One would expect a "stochastic version" of Lipschtz condition as sufficient to show stochastic equicontinuity:

\begin{lem}\label{lem_stochastic_lipschtz}
Let $T(s)$ be defined as in \eqref{eq1_sec3_subsec1}. Suppose for some sequence of random variables $\{B_n\}$, $\|T(s_1)-T(s_2)\|\leq B_n|s_1-s_2|$ holds for all $s_1,s_2\in S$. Then $T(s)$ is stochastic equicontinuous on $S$, if \begin{equation*}
    \lim_{M\rightarrow\infty}\limsup_{n\rightarrow\infty}\prob\lp B_n\geq M\rp=0
\end{equation*} 
\end{lem}

For details, please refer to~\cite{Newey1991}. Lastly, we provide conditions on $T(s)$, $g_i^{-1}(s)$ to ensure such $\{B_n\}$ exists, hence show the uniform convergence of $T(s)$:

\begin{thm}

\label{thm_T_unifm_conv}

Let $T(s)$ and $\bar{g}(s)$ be defined as in \eqref{eq1_sec3_subsec1} and \eqref{eq2_sec3_subsec1}, respectively. Given $S$ a compact subset of $\compl$, and $\bar{g}(s)$ is uniform continuous on $S$. Suppose for some sequence of random variables $\{\Tilde{B}_n\}$, $\max_{i\in[n]}|g_i^{-1}(s_1)-g_i^{-1}(s_2)|\leq \Tilde{B}_n|s_1-s_2|$ holds for all $s_1,s_2\in S$.
Then $T(s)$ uniformly converges in probability to $\frac{1}{n}\bar{g}(s)\one\one^T$
if the following hold:
\begin{enumerate}
    \item $T(s)$ converges point-wise in probability to $\frac{1}{n}\bar{g}\one\one^T$ $\forall s\in S$.
    \item 
\begin{subequations}
\begin{equation}
    \lim_{M\rightarrow\infty}\limsup_{n\rightarrow\infty}\prob\lp \Tilde{B}_n\geq M\rp=0\label{eq_unifm_conv_cond_1}
\end{equation}
\begin{equation}
    \lim_{M\rightarrow\infty}\limsup_{n\rightarrow\infty}\prob\lp \sup_{s\in S}\|T(s)\|\geq M\rp=0\label{eq_unifm_conv_cond_2}
\end{equation}
\end{subequations}
\end{enumerate}
\end{thm}

\ifthenelse{\boolean{archive}}{
\begin{proof}
For all $s_1,s_2\in S$, we have:
\begin{align*}
    &\;\|T(s_1)-T(s_2)\|\\
    =&\;\|T(s_1)(T^{-1}(s_2)-T^{-1}(s_1))T(s_2)\|\\
    \leq & \; \|T(s_1)\|\|T(s_2)\|\|\mathrm{diag}\{g_i^{-1}(s_1)-g_i^{-1}(s_2)\}\|\\
    \leq & \; \lp\sup_{s\in S}\|T(s)\|\rp^2\Tilde{B}_n|s_1-s_2|
\end{align*}
by Lemma \eqref{lem_ineq_sum_prod_bd}, we have:
\begin{align*}
    &\;\prob\lp\lp \sup_{s\in S}\|T(s)\|\rp^2\Tilde{B}_n\geq M\rp\\
    \leq &\; 2\prob \lp\sup_{s\in S}\|T(s)\|\geq M^{1/3} \rp+\prob \lp \Tilde{B}_n\geq M^{1/3} \rp
\end{align*}
Taking $\lim_{M\rightarrow \infty}\limsup_{n\rightarrow \infty}$ on both sides:
\begin{equation*}
    \lim_{M\rightarrow \infty}\limsup_{n\rightarrow \infty}\prob\lp\lp \sup_{s\in S}\|T(s)\|\rp^2\Tilde{B}_n\geq M\rp=0
\end{equation*}
by Lemma \ref{lem_stochastic_lipschtz}, $T(s)$ is stochastic equicontinuous on $S$. Therefore, by Theorem \ref{thm_unifm_conv}, uniform convergence of $T(s)$ on $S$ is proved.
\end{proof}
}{For proof, please refer to~\cite{min2019}.}
\begin{rem}
    For our purpose, the uniform convergence of $T(s)$ over a low frequency band $S=\{jw:w\in[-w_0,w_0]\}$ is sufficient for a stable network to exhibit synchronized output under low frequency disturbance. And as one can expect, the choice of a large enough network size $n_0$ to synchronize the output is frequency dependent, i.e. if we want uniform convergence over a wider frequency band, $n_0$ needs to be larger.   
\end{rem}}{The point-wise convergence in probability suggests that very likely $T(s)$ is "close" to $\frac{1}{n}\bar{g}(s)\one\one^T$ for  large-scale and tightly connected networks. However, for a stable network to exhibit synchronized output under low frequency disturbance, we need to ensure the uniform convergence of $T(s)$ over a low frequency band $S=\{jw:w\in[-w_0,w_0]\}$ for some desired cut-off frequency $w_0$. In general, such  uniform convergence result is not trivial to guarantee.  We refer to \cite{min2019} for conditions that guarantee uniform convergence. Instead, here, we provide numerical evidence that suggests it.}

\section{Numerical verification}\label{sec:numerical}
    
    We now provide two numerical examples of networked dynamics that illustrate the uniform convergence of their transfer matrix over low frequency band by numerical simulation. We will further use our dynamics concentration results propose a reduced order model for power networks.  

    \subsection{Dynamics concentration in consensus networks}
    Consider the network in Fig.\ref{blk_p_n} with the dynamics of each node  given by
    \begin{equation*}
        g_i(s)=\frac{k_i}{s}.
    \end{equation*}
    This gives us the standard continuous time consensus network\cite{Olfati-Saber2007}. Notice that when $g_i(s)$ is simulated by impulse input $\delta (t)$, it is equivalent to setting initial condition at $t=0$ of node $i$ to be $k_i$. 
    
    We sample $k_i$ from $Unif(1,5)$, then $\frac{s}{k_i}$ is sub-Guassian for any fixed $s$. The graph of the network is a $k$-regular ring, i.e. every node is connected to its $2k+1$ nearest neighbors. We set the weight to be $1$ for all edges and $k\approx 0.15n$, then each node is connected to roughly $1/3$ of other nodes in the network. It can be shown that the algebraic connectivity of the graph Laplacian $|\lambda_2|$ is $\Omega(n)$~\cite{Olfati-Saber2007a}, then this network should exhibit dynamics concentration, according to the convergence result in Section \ref{sec:stoc_convergence}. The expected dynamics is given by
    \begin{equation*}
        \bar{g}(s)=\frac{1}{s}\lp \expc\frac{1}{k_i}\rp^{-1}=\frac{4}{\ln 5}\frac{1}{s}
    \end{equation*}
    Suppose the network is subject to an impulse input $u(t)=\delta(t)\one$. Then, the impulse response of $\bar{g}(s)$ is given by $\bar{g}(t)=\frac{4}{\ln 5}\chi_{\geq0}(t)$, where $\chi_{\geq0}(t)$ is the unit step function. 
    
\begin{figure}[h]
        \centering
        \includegraphics[width=7cm]{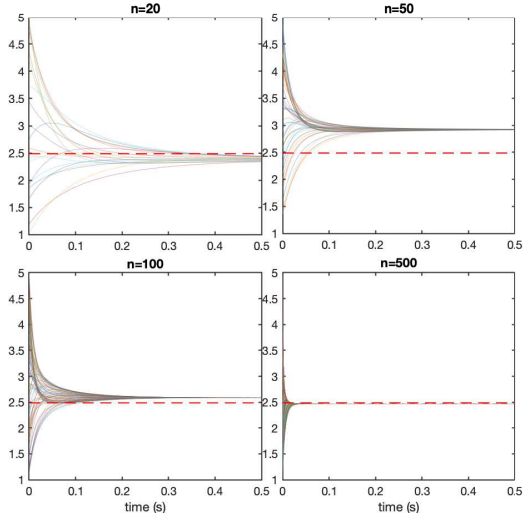}
        \caption{Dynamics Concentration in Consensus network}
        \label{fig_dymC_consens}
    \end{figure}

    We plot the impulse response for network with different size $n=20,50,100,500$ in Fig. \ref{fig_dymC_consens}, along with $\bar g(t)$ shown by the red dashed line.
    It is clear in Fig.~\ref{fig_dymC_consens} that as the network size grows, the network exhibits a more coherent response and the consensus value gets close to the expected average $4/\ln 5$. Although such response can be explained using standard consensus algorithm analysis, it can also be viewed as the consequence of $T(s)$ getting elementwise close to the coherent dynamics $\bar{g}(s)=\frac{4}{n \ln 5}\frac{1}{s}$. 
     
    \subsection{Dynamic concentration in power networks}
     We now look at the case of a power network and leverage our analysis to  provide an accurate low-frequency reduced order model for power networks.
    
    Consider the transfer matrix of power generator networks\cite{Paganini2019} linearized around its equilibrium point, with the following block diagrams:
    
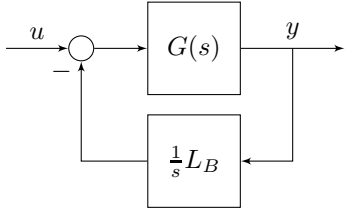
\begin{figure}[ht]
	\centering
	\begin{tikzpicture}[auto, node distance=1.5cm,>=latex']
	\node [input, name=input] {};
	\node [sum, right of=input] (sum) {};
	\node [block, right of=sum] (plant) {$G(s)$};
	\node [output, right of=plant, node distance=2cm] (output) {};
	\node [block, below of=plant] (laplacian) {$\frac{1}{s}L_B$};
	\draw [draw,->] (input) -- node {$u$} (sum);
	\draw [->] (sum) -- (plant);
	\draw [->] (plant) -- node [name=y]{$y$} (output);
	\draw [->] (y) |- (laplacian);
	\draw [->] (laplacian) -| node[pos=0.95]{$-$}(sum);
		
	\end{tikzpicture}
	\caption{Block Diagram of Linearized Power Networks}\label{blk_power}
\end{figure}

    \begin{rem}
        The integration $\frac{1}{s}$ in the feedback loop does not affect the convergence result around $s=0$ because when evaluating $T(s)$ at a particular point $s_0$ around $0$, we notice that $\frac{1}{s_0}$ actually scales up the algebraic connectivity of $\frac{1}{s}L_B$, which makes the network dynamics concentrate to a single dynamic in the consensus subspace. 
    \end{rem}
    The generator dynamics $g_i(s)$ is given by, 
    \begin{equation*}
        g_i(s)=\frac{1}{m_is+d_i}\,,
    \end{equation*}
    for the swing dynamics, where $m_i$ is the inertial and $d_i$ is the damping ratio. 
    
    For generators with turbine control, the transfer function is given by,
    \begin{equation*}
        g_i(s)=\frac{\tau_i s+1}{(m_is+d_i)(\tau_i s+1)+r_i^{-1}}\,,
    \end{equation*}
    where $\tau_i$ is the turbine constant, and $r_i$ is the droop coefficient. We use the Icelandic grid data available at \cite{iceland} where 35 generators in total are connected to the grid, with only some of them implementing turbine control. 
    
    According to our convergence result, for low frequencies or tightly connected networks, the transfer matrix of the network is close to $\frac{1}{n}\Tilde{g}(s)\one\one^T$, with 
    \begin{equation*}
        \Tilde{g}(s)=\lp\frac{1}{n}\sum_{i=1}^ng_i^{-1}(s)\rp^{-1}
    \end{equation*} i.e. we assume $g_i(s)$ are drawn from some unknown distribution and use $\Tilde{g}(s)$ as an empirical approximation of $\bar{g}(s)$. Hence we expect $\Tilde{g}(s)$ to be a good candidate for a reduced model of the whole network.
    
    For generators without turbine control, the reduced model is simply a single generator with its coefficients given by the mean of coefficients among all nodes. However, notice that for generators with turbine control, the reduced model $\Tilde{g}(s)$ is with the same order as a single generator only when the turbine constant $\tau_i$ are the same for all generators in the grid. Every time a different $\tau_i$ is introduced, the order of $\Tilde{g}(s)$ will be increased by $1$. 
    
    In the Icelandic grid, the $\tau_i$ are not the same among generators with turbine control. We will compare the output of:
    \begin{enumerate}
        \item Reduced model $\frac{1}{n}\hat{g}(s)\one\one^T$,  $\hat{g}(s)$ as a representative generator:
        \begin{equation*}
            \hat{g}(s) = \frac{\bar{\tau}s+1}{(\bar{m}s+\bar{d})(\bar{\tau}s+1)+\bar{r}^{-1}}
        \end{equation*}
        with the coefficients:
        \begin{align*}
            &\;\bar{m}=\frac{1}{n}\sum_{i=1}^nm_i,\ \bar{d}=\frac{1}{n}\sum_{i=1}^nd_i\\
            &\;\bar{r}^{-1}=\frac{1}{n}\sum_{i=1}^nr_i^{-1},\ \bar{\tau}=\frac{1}{|\mathcal{T}|}\sum_{i\in\mathcal{T}}\tau_i
        \end{align*}
        where $\mathcal{T}$ is the set of indices such that $\tau_i\neq 0$;
        \item Higher order reduced model $\frac{1}{n}\Tilde{g}(s)\one\one^T$, where
        \begin{equation*}
            \Tilde{g}(s)=\lp\bar{m}s+\bar{d}+\frac{1}{n}\sum_{i\in\mathcal{T}}\frac{r_i^{-1}}{\tau_is+1}\rp^{-1}
        \end{equation*}which is 4th order for Icelandic grid.
    \end{enumerate}
    
    \begin{figure}[h]
        \centering
        \includegraphics[width=7cm]{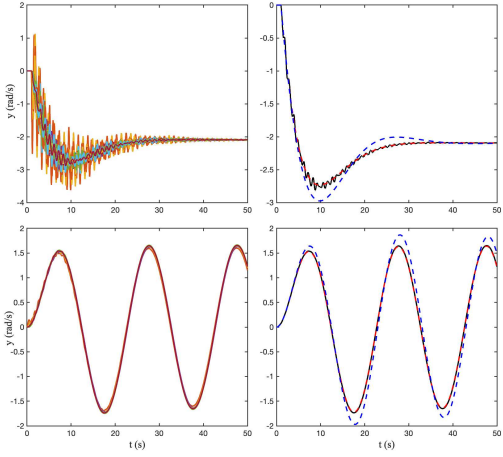}
        \caption{Output response of the power networks and proposed reduced models under step(above) and sinusoidal(below) disturbance. On the left is the output response of each generator. On the right, black solid line: the average output $\bar{y}$ of all generators, red dashed line: output of reduced model $\Tilde{g}(s)$, blue dashed line: output of reduced model $\hat{g}(s)$}
        \label{fig:grid}
    \end{figure}
    
    In Fig.\ref{fig:grid}, we show the output response of the power networks and our proposed reduced model under step $u(t)= -0.3e_2$ and sinusoidal $u(t)=0.2\sin(0.1\pi t)e_2$ disturbance. It is clear that the generators have highly synchronized response to the low frequency disturbance, which shows that even in relatively small networks, the dynamics concentration can be observed in low frequency range, mostly because the integration in the feedback loop scales up $|\lambda_2|$ of the Laplacian, forcing the dynamics of all nodes to concentrate. However, we also observed less synchronized behavior when the network is subject to a step disturbance. This is because at high frequency ranges the integration in the feedback loop scales down $|\lambda_2|$, making the convergence result fail. Consequently, generators have different responses to the high frequency components of the step signal. 
    
    Finally, when comparing the two proposed reduced models, the higher order reduced model is a much better approximation of the original model than a reduced model of a single generator. For networks of generators with turbine control, if $\tau_i$ varies among the network, then $\Tilde{g}(s)$ will have as many poles as the size of the network! Then it is difficult to find a good low order reduced models of such networks as its size scales up.

\section{Conclusion}\label{sec:conclusions}
    
    In this paper, we consider a tightly-connected network consisting of heterogeneous linear node dynamics represented by random transfer functions. We show that the transfer matrix of the network converges in probability as $n$ grows to infinity to a common deterministic scalar transfer function spanning the consensus subspace. We then provide conditions under which such convergence is uniform over a compact set, and numerically verified, for certain networks, that dynamics concentration do occur in the low frequency range. 
    
    There are many possible extension to the current analysis. Firstly, our results suggest that for tightly connected networks, a common controller might perform as good as individual controllers specified for each node, even when nodes are with different types of dynamics. Secondly, another interesting question is whether there are circumstances under which the stability of the concentrated dynamics $\bar g(s)$ imply the stability of the whole network, given large enough network size. It is our belief that this framework can potentially provide new control design tools for large-scale networks.

\ifthenelse{\boolean{archive}}{

\appendix
\def\thesubsubsection{\Alph{section}}

The following Lemmas will be used in the proof of Lemma \ref{lem_hoeff_compl}, Lemma \ref{lem_spec_norm_bd} and Theorem \ref{thm_T_unifm_conv}
\begin{lem}
\label{lem_ineq_sum_prod_bd}
Let $X_1,\cdots,X_M$ be non-negative random variables, then for $\epsilon>0$, we have:
\begin{subequations}
\begin{equation}\label{sum_bd}
    \prob\lp \sum_{i=1}^M X_i\geq \epsilon\rp\leq \sum_{i=1}^M\prob\lp X_i\geq \frac{\epsilon}{M}\rp
\end{equation}
\begin{equation}\label{prod_bd}
    \prob\lp \prod_{i=1}^M X_i\geq \epsilon\rp\leq \sum_{i=1}^M\prob\lp X_i\geq \epsilon^{1/M}\rp
\end{equation}
\end{subequations}
\end{lem}
\begin{proof}
    \begin{align*}
        \prob\lp \sum_{i=1}^M X_i\geq \epsilon\rp&\leq\; \prob\lp\max_{1\leq i\leq M}X_i\geq \frac{\epsilon}{M}\rp\\
        &=\; \prob\lp \bigcup_{i=1}^M\lb X_i\geq \frac{\epsilon}{M}\rb\rp\\
        &\leq\; \sum_{i=1}^M\prob\lp X_i\geq \frac{\epsilon}{M}\rp
    \end{align*}
    Similarly, since $X_i\geq 0$
    \begin{align*}
        \prob\lp \prod_{i=1}^M X_i\geq \epsilon\rp&\leq\; \prob\lp\max_{1\leq i\leq M}X_i\geq \epsilon^{1/M}\rp\\
        &\leq\; \sum_{i=1}^M\prob\lp X_i\geq \epsilon^{1/M}\rp
    \end{align*}
\end{proof}
\begin{lem}
\label{lem_ineq_recip_bd}
Let $X,Y,Z,W_1,W_2$ be non-negative random variables. Suppose whenever $Y\neq 0$, $\frac{1}{Y}\geq W_1-W_2$ holds, then we have:
\begin{equation*}
    \prob\lp X Y\geq Z\rp\leq \prob\lp W_2\geq W_1\rp +\prob\lp X\geq Z(W_1-W_2)\rp
\end{equation*}
\begin{proof}
    when $W_2<W_1$, by $\frac{1}{Y}\geq W_1-W_2$, we obtain:
    \begin{equation*}
        Y\leq \frac{1}{W_1-W_2}
    \end{equation*}
    Notice that the inequality above also holds for $Y=0$ as long as $W_2<W_1$, hence,
    \begin{align*}
        \prob\lp XY\geq Z\rp&=\;\prob((\{ W_2\geq W_1\}\cap\{XY\geq Z\})\cup\\
        &\; \qquad\  (\{ W_2< W_1\}\cap\{XY\geq Z\}))\\
        &\leq\; \prob(\{ W_2\geq W_1\}\cap\{XY\geq Z\})+\\
        &\;\qquad\ \prob((\{ W_2< W_1\}\cap\{XY\geq Z\})\\
        &\leq\; \prob(W_2\geq W_1)+\\
        &\;\qquad\ \prob\lp \{W_2< W_1\}\cap \lb \frac{X}{W_1-W_2}\geq Z\rb\rp\\
        &=\; \prob(W_2\geq W_1)+\prob\lp X\geq Z(W_1-W_2)\rp
    \end{align*}
\end{proof}
\end{lem}

\begin{proof}[Proof of Lemma \ref{lem_hoeff_compl}]
Firstly, we show that $\forall t>0$
\begin{align*}
    &\;\prob\lp \lv\sum_{i=1}^{n}a_iX_i\rv \geq t\rp\\
    \leq &\;\prob\lp \lv\sum_{i=1}^{n}Re(a_i)X_i\rv+\lv \sum_{i=1}^{n}Im(a_i)X_i\rv \geq t\rp\\
    &\; \text{by Lemma \ref{lem_ineq_sum_prod_bd}}\\
    \leq &\;\prob\lp \lv\sum_{i=1}^{n}Re(a_i)X_i\rv\geq \frac{t}{2}\rp+\prob\lp \lv\sum_{i=1}^{n}Im(a_i)X_i\rv\geq \frac{t}{2}\rp\\
    &\; \text{by Lemma \ref{hoeff_ineq}, for some } c_x>0:\\
    \leq &\;2\exp\lp -\frac{c_xt^2}{4\sum_{i=1}^n Re^2(a_i)}\rp+2\exp\lp -\frac{c_xt^2}{4\sum_{i=1}^n Im^2(a_i)}\rp\\
    \leq &\;2\exp\lp -\frac{c_xt^2}{4\|a\|^2}\rp+2\exp\lp -\frac{c_xt^2}{4\|a\|^2}\rp=4\exp\lp -\frac{c_xt^2}{4\|a\|^2}\rp
\end{align*}
Similarly, for some $c_y>0$:
\begin{equation*}
    \prob\lp \lv\sum_{i=1}^{n}a_iY_i\rv \geq t\rp\leq 4\exp\lp -\frac{c_yt^2}{4\|a\|^2}\rp
\end{equation*}
Then we have:
\begin{align*}
    &\;\prob\lp \lv\sum_{i=1}^{n}a_i(X_i+jY_i)\rv \geq t\rp\\
    \leq &\; \prob\lp \lv\sum_{i=1}^{n}a_iX_i\rv+\lv \sum_{i=1}^{n}a_iY_i\rv \geq t\rp\\
    &\; \text{by Lemma \ref{lem_ineq_sum_prod_bd}}\\
    \leq &\;\prob\lp \lv\sum_{i=1}^{n}a_iX_i\rv\geq \frac{t}{2}\rp+\prob\lp \lv\sum_{i=1}^{n}a_iY_i\rv\geq \frac{t}{2}\rp\\
    &\; \text{apply the bounds we obtained above:}\\
    \leq &\; 4\exp\lp -\frac{c_xt^2}{16\|a\|^2}\rp+4\exp\lp -\frac{c_yt^2}{16\|a\|^2}\rp\\
    \leq &\; 8\exp\lp -\frac{ct^2}{\|a\|^2}\rp,\quad \text{where }c=\min\lb\frac{c_x}{16},\frac{c_y}{16}\rb
\end{align*}
\end{proof}

\begin{proof}[Proof of Lemma \ref{lem_concenration}]
Firstly, for every $k\in[n]$ we have $d_{k1}=\sum_{i=1}^{n}\bar{v}_{ki}v_{1i}(\Tilde{X}_i+j\Tilde{Y}_i)$, by Lemma \ref{lem_hoeff_compl}, for $\epsilon>0$ we have \begin{equation*}
    \prob\lp \lv \sum_{i=1}^{n}\bar{v}_{ki}v_{1i}(\Tilde{X}_i+j\Tilde{Y}_i)\rv\geq \epsilon\rp\leq 8\exp\lp -\frac{c_1\epsilon^2}{\sum_{i=1}^{n}|\bar{v}_{ki}v_{1i}|^2}\rp
\end{equation*}
for some $c_1>0$. Notice that $v_1=\one/\sqrt{n}$, then\[
\sum_{i=1}^{n}|\bar{v}_{ki}v_{1i}|^2=\frac{1}{n}\sum_{i=1}^{n}|\bar{v}_{ki}|^2=\frac{1}{n}\]
Then we have\begin{equation*}
    \prob\lp |d_{k1}|\geq \epsilon\rp\leq 8\exp\lp -c_1n\epsilon^2\rp,\quad \forall k\in[n]
\end{equation*}
which proves the firstly inequality. Similarly, for $d_{1l}=\sum_{i=1}^{n}\bar{v}_{1i}v_{li}(\Tilde{X}_i+j\Tilde{Y}_i)$, $l\in[n]$, we have:
\begin{equation*}
    \prob\lp |d_{1l}|\geq \epsilon\rp\leq 8\exp\lp -c_1n\epsilon^2\rp,\quad \forall l\in[n]
\end{equation*}
With the inequality above, the 2nd moment of $d_{k1}$ is bounded by:
\begin{align*}
    \expc|d_{k1}|^2&=\;\int_0^{+\infty}2t\prob\lp|d_{k1}|\geq t\rp dt\\
    &\leq \;8\int_0^{+\infty}2t\exp\lp-c_1 nt^2\rp dt = \frac{8}{c_1 n}:=\frac{c_2}{n}
\end{align*}
From this, by Markov's inequality, we have:
\begin{align*}
    \prob\lp\sum_{k=1}^n|d_{k1}|^2\geq \epsilon n^p\rp&\leq\; \frac{\sum_{k=1}^n\expc|d_{k1}|^2}{\epsilon n^p}\leq \frac{c_2}{\epsilon n^p}
\end{align*}
which gives inequality \eqref{lem_conc_eq2_sub1}, similarly for inequality \eqref{lem_conc_eq2_sub2}.

Now for the last inequality, since unitary matrix preserves the spectral norm, it is straightforward to see that
\begin{align*}
\|D\|&=\;\|V^*\mathrm{diag}\{\Tilde{X}_i+j\Tilde{Y}_i\}V\|\\
&=\;\|\mathrm{diag}\{\Tilde{X}_i+j\Tilde{Y}_i\}\|=\max_{i\in[n]}|\Tilde{X}_i+j\Tilde{Y}_i|    
\end{align*}
We show that $|\Tilde{X}_i+j\Tilde{Y}_i|$ is also sub-gaussian by noticing that:
\begin{equation*}
    \prob\lp |\Tilde{X}_i+j\Tilde{Y}_i|>t\rp\leq \prob\lp |\Tilde{X}_i|+|\Tilde{Y}_i|>t\rp\nonumber\
\end{equation*}
and $|X_i|+|Y_i|$ is a sum of two sub-gaussian random variables, hence sub-gaussian~\cite[Proposition 2.6.1]{Vershynin2018}. For sequence of sub-gaussian random variables, its expected maximum is bounded by~\cite[Excercise 2.5.10]{Vershynin2018}:
\begin{equation*}
    \expc \max_{i\in[n]}|\Tilde{X}_i+j\Tilde{Y}_i|\leq c_3\sqrt{\log n}
\end{equation*}
for some $c_3>0$. Then we have:
\begin{align*}
    \prob\lp\|D\|\geq n^p\epsilon\rp&=\; \prob\lp\max_{i\in[n]}|\Tilde{X}_i+\Tilde{Y}_i|\geq n^{p}\epsilon\rp\\
    (\text{Markov's Inequality})\ &\leq\; \frac{\expc \max_{i\in[n]}|\Tilde{X}_i+j\Tilde{Y}_i|}{n^{p}\epsilon}\leq \frac{c_3\sqrt{\log n}}{n^{p}\epsilon}
\end{align*}
\end{proof}

\begin{proof}[Proof of Lemma \ref{lem_spec_norm_bd}]
Firstly, notice that $\|H^{-1}-\mu^{-1} e_1e_1^T\|$ is bounded by the summation of the spectral norm of every blocks of $H^{-1}$:
\begin{align}
    &\;\|H^{-1}-\mu^{-1} e_1e_1^T\|\nonumber\\ \leq &\;\lv\frac{1}{a}-\frac{1}{\mu}\rv+\lV\frac{1}{a}h_{12}^TH_{22}^{-1}\rV+\lV\frac{1}{a}H_{22}^{-1}h_{21}\rV\nonumber\\
    &\;\qquad\quad +\lV H_{22}^{-1}+\frac{1}{a}H_{22}^{-1}h_{21}h_{12}^TH_{22}^{-1} \rV\nonumber\\
   \leq &\; \lv\frac{1}{a}-\frac{1}{\mu}\rv+\lV\frac{1}{a}h_{12}^TH_{22}^{-1}\rV+\lV\frac{1}{a}H_{22}^{-1}h_{21}\rV\nonumber\\
    &\;\qquad\quad +\lV H_{22}^{-1}\rV+\lV\frac{1}{a}H_{22}^{-1}h_{21}h_{12}^TH_{22}^{-1} \rV\label{eq_norm_bd_split}
\end{align}
Then by Lemma \eqref{lem_ineq_sum_prod_bd}, we have:
\begin{align}
    &\;\prob\lp\lV H^{-1}-\mu^{-1}e_1e_1^T\rV\geq\epsilon\rp\nonumber\\
    \leq &\; \prob\lp\lv\frac{1}{a}-\frac{1}{\mu}\rv\geq \frac{\epsilon}{5}\rp+\prob\lp\|H_{22}^{-1}\|\geq \frac{\epsilon}{5}\rp\nonumber\\
    &\;\ +\prob\lp \lV\frac{1}{a}h_{12}^TH_{22}^{-1}\rV\geq \frac{\epsilon}{5}\rp +\prob\lp \lV\frac{1}{a}H_{22}^{-1}h_{21}\rV\geq \frac{\epsilon}{5}\rp\nonumber\\
    &\;\ +\prob\lp \lV\frac{1}{a}H_{22}^{-1}h_{21}h_{12}^TH_{22}^{-1} \rV\geq \frac{\epsilon}{5}\rp\label{tail_prob_split}
\end{align}
The remaining proof is to bound the tail probabilities above.

Notice that $\|H_{22}^{-1}\|=\frac{1}{\underaccent{\bar}{\sigma}(H_{22})}$
By Weyl's inequality~\cite[Theorem \MakeUppercase{\romannumeral 3}.2.1]{Bhatia2013}:
\begin{equation}
    |\underaccent{\bar}{\sigma}(H_{22})-\underaccent{\bar}{\sigma}(H_{22}-D_{(1)(1)})|\leq \|D_{(1)(1)}\|
\end{equation}
Then we know that: 
\begin{align*}
    \frac{1}{\|H_{22}^{-1}\|}=\underaccent{\bar}{\sigma}(H_{22})&\geq\; \underaccent{\bar}{\sigma}(H_{22}-D_{(1)(1)})-\|D_{(1)(1)}\|\\
    &=\; \min_{2\leq i\leq n}|\lambda_i+\mu|-\|D_{(1)(1)}\|\\
    &\geq\; \min_{2\leq i\leq n}|\lambda_i|-|\mu|-\|D_{(1)(1)}\|\\
    &=\; |\lambda_2|-|\mu|-\|D_{(1)(1)}\|
\end{align*}
Since $|\lambda_2|=\Omega(n^p)$, then $\exists m_1>0$ s.t. for some $n_1>0$, whenever $n\geq n_1$, $|\lambda_2|-|\mu|\geq m_1n^p$ holds, which leads to:
\begin{equation}
    \frac{1}{\|H_{22}^{-1}\|}\geq m_1n^p-\|D_{(1)(1)} \|,\ n\geq n_1  \label{eq_h22_inv_bd}
\end{equation}
For the bound on tail probability of $\|H_{22}^{-1}\|$, by Lemma \ref{lem_ineq_recip_bd}:
\begin{align}
    &\;\prob\lp \|H_{22}^{-1}\|\geq \frac{\epsilon}{5}\rp\nonumber\\
    \leq &\; \prob\lp\|D_{(1)(1)}\|\geq m_1n^p\rp+\prob\lp 1\geq \frac{\epsilon}{5}m_1n^p-\frac{\epsilon}{5}\|D_{(1)(1)}\|\rp\nonumber\\
    = &\; \prob\lp\|D_{(1)(1)}\|\geq m_1n^p\rp+\prob\lp \|D_{(1)(1)}\|\geq m_1n^p-\frac{5}{\epsilon}\rp\nonumber\\
    \leq &\; 2\prob\lp \|D_{(1)(1)}\|\geq m_1n^p-\frac{5}{\epsilon}\rp
\end{align}
Because $\exists 0<m_2<m_1$ s.t. when $n\geq n_2$, $m_1n^p-\frac{5}{\epsilon}\geq m_2n^p$ holds, we let $n\geq \max\{n_1,n_2\}$, to get:
\begin{align}
    \prob\lp \|H_{22}^{-1}\|\geq \frac{\epsilon}{5}\rp&\leq\; 2\prob\lp \|D_{(1)(1)}\|\geq m_1n^p-\frac{5}{\epsilon}\rp\nonumber\\
    &\leq\; 2\prob\lp \|D_{(1)(1)}\|\geq m_2n^p\rp\label{eq_reduc_0}
\end{align}
Now we turn to bound the relatively complex terms:
\begin{align}
    &\;\prob\lp\lV\frac{1}{a}h_{12}^TH_{22}^{-1}\rV\geq \frac{\epsilon}{5}\rp\nonumber\\
    = &\; \prob\lp\lv\frac{1}{a}-\frac{1}{\mu}+\frac{1}{\mu}\rv\lV h_{12}^TH_{22}^{-1}\rV\geq \frac{\epsilon}{5}\rp\nonumber\\
    \leq &\; \prob\lp\lp\lv\frac{1}{a}-\frac{1}{\mu}\rv+\lv\frac{1}{\mu}\rv\rp\lV h_{12}^TH_{22}^{-1}\rV\geq \frac{\epsilon}{5}\rp\nonumber\\
    &\;\ \text{by Lemma \ref{lem_ineq_sum_prod_bd}}\nonumber\\
    \leq &\; \prob\lp\lv\frac{1}{a}-\frac{1}{\mu}\rv\lV h_{12}^TH_{22}^{-1}\rV\geq \frac{\epsilon}{10}\rp+\prob\lp\lV h_{12}^TH_{22}^{-1}\rV\geq \frac{\epsilon |\mu|}{10}\rp\nonumber\\
    &\;\ \text{apply Lemma \ref{lem_ineq_sum_prod_bd} again on first term}\nonumber\\
    \leq &\; \prob\lp\lv\frac{1}{a}-\frac{1}{\mu}\rv\geq \sqrt{\frac{\epsilon}{10}}\rp+ \prob\lp\lV h_{12}^TH_{22}^{-1}\rV\geq \sqrt{\frac{\epsilon}{10}}\rp\nonumber\\
    &\;\quad +\prob\lp\lV h_{12}^TH_{22}^{-1}\rV\geq \frac{\epsilon |\mu|}{10}\rp\nonumber\\
    \leq &\;\prob\lp\lv\frac{1}{a}-\frac{1}{\mu}\rv\geq \sqrt{\frac{\epsilon}{10}}\rp+2\prob\lp\lV h_{12}^TH_{22}^{-1}\rV\geq \epsilon_1\rp\label{eq_reduc_1}
\end{align}
where $\epsilon_1=\min\{\frac{\epsilon |\mu|}{10},\sqrt{\frac{\epsilon}{10}}\}$.\\
Similarly, we have:
\begin{align}
    &\;\prob\lp\lV\frac{1}{a}H_{22}^{-1}h_{21}\rV\geq \frac{\epsilon}{5}\rp\nonumber\\
    \leq &\;\prob\lp\lv\frac{1}{a}-\frac{1}{\mu}\rv\geq \sqrt{\frac{\epsilon}{10}}\rp+2\prob\lp\lV H_{22}^{-1}h_{21}\rV\geq \epsilon_1\rp\label{eq_reduc_2}
\end{align}
and
\begin{align}
    &\;\prob\lp\lV\frac{1}{a}H_{22}^{-1}h_{21}h_{12}^TH_{22}^{-1}\rV\geq \frac{\epsilon}{5}\rp\nonumber\\
    \leq &\;\prob\lp\lv\frac{1}{a}-\frac{1}{\mu}\rv\geq \sqrt{\frac{\epsilon}{10}}\rp+2\prob\lp\lV H_{22}^{-1}h_{21}h_{12}^TH_{22}^{-1}\rV\geq \epsilon_1\rp\nonumber\\
    \leq &\;\prob\lp\lv\frac{1}{a}-\frac{1}{\mu}\rv\geq \sqrt{\frac{\epsilon}{10}}\rp+2\prob\lp\lV H_{22}^{-1}h_{21}\rV\lV h_{12}^TH_{22}^{-1}\rV\geq \epsilon_1\rp\nonumber\\
    &\;\ \text{apply Lemma \ref{lem_ineq_sum_prod_bd} on second term}\nonumber\\
    \leq &\; \prob\lp\lv\frac{1}{a}-\frac{1}{\mu}\rv\geq \sqrt{\frac{\epsilon}{10}}\rp\nonumber\\
    &\; +2\prob\lp\lV h_{12}^TH_{22}^{-1}\rV\geq \sqrt{\epsilon_1}\rp+2\prob\lp\lV H_{22}^{-1}h_{21}\rV\geq \sqrt{\epsilon_1}\rp \label{eq_reduc_3}
\end{align}
Then we continue to reduce the tail probability of $\lv \frac{1}{a}-\frac{1}{\mu}\rv$, notice that
\begin{equation*}
    |a|=|d_{11}+\mu-h_{12}^TH_{22}^{-1}h_{21}|\geq |\mu|-|d_{11}-h_{12}H_{22}^{-1}h_{21}|
\end{equation*}
then we let $\epsilon_2=\min\{\sqrt{\frac{\epsilon}{10}},\frac{\epsilon}{5}\}$
\begin{align}
    &\;\prob\lp\lv\frac{1}{a}-\frac{1}{\mu}\rv\geq \epsilon_2\rp\nonumber\\
    = &\; \prob\lp \frac{|d_{11}-h_{12}^TH_{22}^{-1}h_{21}|}{|a|}\geq \epsilon_2|\mu|\rp\nonumber\\
     &\;\ \text{by Lemma \ref{lem_ineq_recip_bd} } \nonumber\\
    \leq &\;\prob\lp |d_{11}-h_{12}^TH_{22}^{-1}h_{21}|\geq |\mu|\rp \nonumber\\
    &\;\ + \prob\lp |d_{11}-h_{12}^TH_{22}^{-1}h_{21}|\geq \frac{\epsilon_2|\mu|^2}{1+\epsilon_2 |\mu|}\rp\nonumber\\
    &\;\ \text{notice that }|\mu|\geq \frac{\epsilon_2 |\mu|^2}{1+\epsilon_2|\mu|}\nonumber\\
    \leq &\;2\prob\lp |d_{11}-h_{12}^TH_{22}^{-1}h_{21}|\geq \epsilon_3\rp\nonumber\\
    \leq &\;2\prob\lp |d_{11}|\geq \frac{\epsilon_3}{2}\rp+2\prob\lp |h_{12}^TH_{22}^{-1}h_{21}|\geq \frac{\epsilon_3}{2}\rp\label{eq_reduc_4}
\end{align}
where $\epsilon_3=\frac{\epsilon_2 |\mu|^2}{1+\epsilon_2 |\mu|}$. Apparently, we have:
\begin{subequations}
    \begin{equation}
        \prob\lp\lv\frac{1}{a}-\frac{1}{\mu}\rv\geq \frac{\epsilon}{5}\rp\leq \prob\lp\lv\frac{1}{a}-\frac{1}{\mu}\rv\geq \epsilon_2\rp
    \end{equation}
    \begin{equation}
        \prob\lp\lv\frac{1}{a}-\frac{1}{\mu}\rv\geq \sqrt{\frac{\epsilon}{10}}\rp\leq \prob\lp\lv\frac{1}{a}-\frac{1}{\mu}\rv\geq \epsilon_2\rp
    \end{equation}
\end{subequations}
At last, we bound the tail probabilities of the terms containing $H_{22}^{-1}$ by:
\begin{align}
    &\;\prob\lp |h_{12}^TH_{22}^{-1}h_{21}|\geq \frac{\epsilon_3}{2}\rp\nonumber\\
    \leq &\; \prob\lp 2\|h_{12}\|\|H_{22}^{-1}\|\|h_{21}\|\geq \epsilon_3\rp\nonumber\\
    &\;\ \text{by Lemma \ref{lem_ineq_recip_bd}}\nonumber\\
    \leq &\; \prob\lp \|D_{(1)(1)}\|\geq m_1n^p\rp+\nonumber\\
    &\;\quad \prob\lp2\|h_{12}\|\|h_{21}\|\geq \epsilon_3 m_1n^p-\epsilon_3 D_{(1)(1)}\rp\nonumber\\
    \leq &\; \prob\lp \|D_{(1)(1)}\|\geq m_1n^p\rp+\nonumber\\
    &\;\quad \prob\lp2\|h_{12}\|\|h_{21}\|+\epsilon_3 D_{(1)(1)}\geq \epsilon_3 m_1n^p\rp\nonumber\\
    &\;\ \text{Apply Lemma \ref{lem_ineq_recip_bd} twice on second term}\nonumber\\
    \leq &\; \prob\lp \|D_{(1)(1)}\|\geq m_1n^p\rp+\nonumber\\
    &\;\quad \prob\lp2\|h_{12}\|\|h_{21}\|\geq \frac{\epsilon_3}{2} m_1n^p\rp+\prob\lp \|D_{(1)(1)}\|\geq \frac{1}{2} m_1n^p\rp\nonumber\\
    \leq &\; 2\prob\lp \|D_{(1)(1)}\|\geq \frac{1}{2}m_1n^p\rp+\nonumber\\
    &\;\quad \prob\lp\|h_{12}\|\geq \frac{\sqrt{\epsilon_3m_1}}{2} n^{p/2}\rp+\prob\lp\|h_{21}\|\geq \frac{\sqrt{\epsilon_3m_1}}{2} n^{p/2}\rp\label{eq_reduc_5}
\end{align}
Similarly, we have:
\begin{align}
    &\;\prob\lp \|h_{12}^TH_{22}^{-1}\|\geq \epsilon_1\rp\nonumber\\ 
    \leq &\; 2\prob\lp \|D_{(1)(1)}\|\geq \frac{1}{2}m_1n^p\rp+\prob\lp \|h_{12}\|\geq \frac{\epsilon_1}{2}m_1n^{p}\rp\label{eq_reduc_6}
\end{align}
and
\begin{align}
    &\;\prob\lp \|H_{22}^{-1}h_{21}\|\geq \epsilon_1\rp\nonumber\\ 
    \leq &\; 2\prob\lp \|D_{(1)(1)}\|\geq \frac{1}{2}m_1n^p\rp+\prob\lp \|h_{21}\|\geq \frac{\epsilon_1}{2}m_1n^{p}\rp\label{eq_reduc_7}
\end{align}
Bounds on $\prob\lp \|h_{12}^TH_{22}^{-1}\|\geq \sqrt{\epsilon_1}\rp$ and $\prob\lp \|H_{22}^{-1}h_{21}\|\geq \sqrt{\epsilon_1}\rp$ can be easily derived from \eqref{eq_reduc_6} and \eqref{eq_reduc_7} respectively. But notice that $\exists n_3>0$ s.t. when $n\geq n_3$, $\frac{\sqrt{\epsilon_3m_1}}{2}n^{p/2}\leq \frac{1}{2}\min\{\epsilon_1,\sqrt{\epsilon_1}\}m_1n^p$, hence the tail probabilities of $\|h_{12}\|$ and $\|h_{21}\|$ in \eqref{eq_reduc_6}\eqref{eq_reduc_7} are always bounded by those in \eqref{eq_reduc_5} when $n$ large enough.   

Combining the bounds in \eqref{eq_reduc_0}-\eqref{eq_reduc_7}, when $n\geq \max\{n_1,n_2,n_3\}$, we have the following:
\begin{align}
    &\;\prob\lp\|H^{-1}-\mu^{-1}e_1e_1^T\|\geq \epsilon\rp\nonumber\\
    \leq &\; 8\prob\lp |d_{11}|\geq C_1\rp+34\prob\lp\|D_{(1)(1)}\|\geq C_2n^p\rp\nonumber\\
    &\;\quad + 12\prob\lp \|h_{12}\|\geq C_3n^{p/2}\rp + 12\prob\lp \|h_{21}\|\geq C_3n^{p/2}\rp\label{eq_tail_split}
\end{align}
where
\begin{align*}
    C_1= \frac{\epsilon_3}{2};\ C_2=\min\{\frac{m_1}{2},m_2\};\ C_3= \frac{\sqrt{\epsilon_3m_1}}{2}
\end{align*}
Lastly, notice that $\|h_{12}\|^2\leq \sum_{l=1}^n|d_{1l}|^2$, $\|h_{21}\|^2\leq \sum_{k=1}^n|d_{k1}|^2$ and $\|D_{(1)(1)}\|\leq\|D\|$, by applying the following bounds:
\begin{align*}
    \prob\lp \|h_{12}\|\geq C_3n^{p/2}\rp&=\;\prob\lp \|h_{12}\|^2\geq C_3^2n^{p}\rp\\
    &\leq \; \prob\lp \sum_{l=1}^n|d_{1l}|^2\geq C_3^2n^{p}\rp\\
    \prob\lp \|h_{21}\|\geq C_3n^{p/2}\rp&\leq\; \prob\lp \sum_{k=1}^n|d_{k1}|^2\geq C_3^2n^{p}\rp\\
    \prob\lp\|D_{(1)(1)}\|\geq C_2n^p\rp&\leq\;\prob\lp\|D\|\geq C_2n^p\rp
\end{align*}
to \eqref{eq_tail_split}, we finishes the proof.
\end{proof}

}{}
\ifthenelse{\boolean{archive}}{}{\balance}
\bibliographystyle{IEEEtran}
\bibliography{ref.bib}
\end{document}